\documentstyle[aps,psfig]{revtex}

\newcommand{\be}{\begin{equation}}
\newcommand{\ee}{\end{equation}}
\newcommand{\bea}{\begin{eqnarray}}
\newcommand{\eea}{\end{eqnarray}}
\newcommand{\bml}{\begin{mathletters}}
\newcommand{\eml}{\end{mathletters}}

\begin{document}
\title{Time Delay Effects on Coupled Limit Cycle Oscillators at Hopf Bifurcation}
\author{D.V. Ramana Reddy,\cite{tapovan} A. Sen and G. L. Johnston\cite{GL:address}}
\address{Institute for Plasma Research, Bhat, Gandhinagar 382 428, India}
\maketitle

\begin{abstract}
We present a detailed study of the effect of time delay on the
collective dynamics of coupled limit cycle oscillators at Hopf
bifurcation. For a simple model consisting of just two
oscillators with a time delayed coupling, the bifurcation
diagram obtained by numerical and analytical solutions shows
significant changes in the stability boundaries of the amplitude
death, phase locked and incoherent regions.  A novel result is
the occurrence of amplitude death even in the absence of a
frequency mismatch between the two oscillators.  Similar results
are obtained for an array of N  oscillators with a delayed mean
field coupling and the regions of such amplitude death in the
parameter space of the coupling strength and time delay are
quantified. Some general analytic results for the $N \rightarrow
\infty$ (thermodynamic) limit are also obtained and the
implications of the time delay effects for physical applications
are discussed.


\end{abstract}

\section{Introduction}
\label{SEC:intro}

The collective behavior of a large assembly of coupled
nonlinear oscillators provides valuable clues for understanding
the complex dynamics of systems with many degrees of freedom.
This has been one of the major motivations for the recent large
scale interest both experimentally\cite{BL:96} and 
numerically\cite{Sat:89} in the study of such simple mathematical 
models and their application to a wide variety of physical and
biological problems.
Examples of such applications include 
interactions of
arrays of Josephson junctions\cite{HBW:88,WCS:96},
semiconductor
lasers\cite{VGEKL:97,HGEK:97}, charge density
waves\cite{GZ:85}, phase locking of relativistic
magnetrons\cite{BSWSH:89}, Belousov-Zhabotinskii reactions in
coupled Brusselator models\cite{SM:82,Kur:84,CE:89,DE:96}
and neural oscillator models for  circadian
pacemakers\cite{KS:80}. One of the prominent cooperative
phenomena, that was first highlighted by Winfree \cite{Win:80} in a simple
model of weakly coupled limit cycle oscillators, is that of
frequency entrainment or synchronization of the diverse
frequencies of the oscillator assembly to a single common
frequency \cite{Kur:84a,Dai:90}.
This happens in a spontaneous and abrupt fashion
once the coupling strength exceeds a critical
threshold. Real life examples of such behavior abound 
in nature e.g. the synchronous flashing of
a swarm of fire flies, the chirping of crickets in unison or the
electrical synchrony in cardiac cells. When the coupling between
the oscillators is comparable to the attraction to their
individual limit cycles, other interesting phenomena 
can occur\cite{Aiz:76,SF:89,PM:92,RW:96}
which involve the amplitudes of the individual oscillators.
For example, if the coupling is sufficiently strong and the spread
in the natural frequencies of the oscillators sufficiently broad,
the oscillators can suffer an amplitude quenching 
or death\cite{AEK:90,Erm:90,MS:90a}. Such
behavior has been observed in experiments of coupled chemical
oscillator systems e.g. coupled Belousov-Zhabotinskii reactions
carried out in coupled tank reactors\cite{Bar:85}. 
Other collective phenomena that these coupled oscillator models display 
include partial synchronization, phase trapping, large amplitude Hopf
oscillations and even chaotic behavior\cite{MS:90,MMS:91} - all of which have been
discussed widely in the literature.\\

The goal of our present study is to examine the effect of time
delay on the collective dynamics of coupled oscillator systems.
Time delay is ubiquitous in most physical and biological
systems like optical bistable devices\cite{VDCD:87},
electromechanical systems\cite{Min:48},
predator-prey models\cite{FV:97}, and physiological systems 
\cite{EPG:95,CW:98}.
They can arise from finite propagation speeds of
signals, for example, or from finite processing times in
synapses, finite reaction times in chemical processes and so on.
Surprisingly most past studies on coupled oscillator systems
have not considered the effect of time delay. The
work of Schuster and Wagner\cite{SW:89}, 
Niebur {\it et al.}\cite{NSK:91}, Nakamura, {\em et al.}\cite{NTM:94}
and Kim, {\em et al.}\cite{KPR:97}, are the only
ones, that we are aware of, where they have tried to incorporate
time delay effects in the context of the coupled oscillator problem.
However, they have restricted themselves to the simplest of
models, that of coupled limit cycle oscillators in the weak
coupling limit where only the phase information is retained and
phenomena like amplitude death cannot occur. Our aim is to
extend this study into the strong coupling regime where both
phase and amplitude responses need to be retained and to investigate
the effect of time delay on the various collective responses of
such a model system.  To keep the analysis simple we begin with
just two limit cycle oscillators that are coupled with a time
delay. For such a model the bifurcation diagram is easy to
obtain both analytically as well as numerically. Furthermore it
allows us a detailed comparison with the past work
of Aronson {\it et al.}\cite{AEK:90} who have analyzed an analogous 
model but without any time delay in the coupling.  After identifying and
briefly discussing the results of time delay effects in this simple
$N=2$ model we next proceed to analyze a large assembly of
coupled limit cycle oscillators. For this study, we construct
a model which is a generalization of the mean
field model of Matthews {\it et al.}\cite{MS:90,MMS:91} and where the coupling term
is suitably modified to introduce time delay. This model is
explored in detail by extensive numerical solutions and linear
stability analysis around fixed points. The limit of 
$N \rightarrow \infty$ is particularly interesting and permits some explicit
analytic results. Our principal focus is on cooperative
phenomena like amplitude death and frequency locking and we find
that both these states are significantly influenced 
by time delay effects.  One of the surprising and dramatic
results is that in the presence of time delay amplitude death
can occur even for zero frequency mismatch between the
oscillators (i.e.  for identical oscillators). This is in sharp
contrast to the situation with no time delay where all previous
numerical and analytical works have emphasized the need to have a
broad frequency spread for amplitude death to occur. A brief
report of this result has been published by us elsewhere\cite{RSJ:98}.
In this paper we give a more detailed and complete description of
this phenomenon. We also report on other newer findings 
related to time delay induced
effects in the collective regimes corresponding to phase locked and 
chaotic states.\\

The organization of the paper is as follows. In the next section (Section \ref{SEC:n2}) 
we analyze the model of two limit cycle oscillators
that have a time delayed coupling and compare and contrast our results
with the previous work of Aronson {\it et al}. In Section \ref{SEC:ngen}, we
describe the more general  $N$ coupled oscillator model and
present numerical as well as analytic results for the
different collective states. This includes the amplitude death region,
the phase locked region and the so called chaotic regime. Some explicit
results for the $N \rightarrow \infty$ (thermodynamic) limit are also
presented. Section \ref{SEC:summary} provides a summary of our results and a brief
discussion on possible future extensions of this work.
\section{Two delay coupled oscillators}
\label{SEC:n2}
\subsection{Model Equations}
\label{SUBSEC:n2model}
For the fundamental oscillator unit of our model we choose the
simple limit cycle oscillator
described by the equation,
\be
\dot{Z}_{j}(t) = (1+{\em i} \omega_{j} - \mid Z_{j}(t) \mid^2) Z_{j}(t)
\ee
where $Z_{j}$ is the complex amplitude of the $jth$ oscillator. Each
oscillator has a stable limit cycle of unit amplitude $\mid Z_{j}\mid ~=~ 1$
with angular frequency $\omega_{j}$. We consider a simple model in which
two of them are coupled linearly to each other as follows,

\bml
\bea
\label{EQN:Z1dot}
\dot{Z}_1(t) & ~=~ & (1+i\omega_1-\mid Z_1(t)\mid^2)~Z_1(t) 
+~K ~[Z_2(t-\tau )~-~Z_1(t)], \\
\label{EQN:Z2dot}
\dot{Z}_2(t) & ~=~ & (1+i\omega_2-\mid Z_2(t)\mid^2)~Z_2(t)
+~K ~[Z_1(t-\tau )~-~Z_2(t)],
\eea
\eml
where $K \ge 0$ is the coupling strength and $\tau \ge 0$ is a measure of the
time delay. These model equations are a
direct generalization of the set of 
equations studied by Aronson, {\it et al.}\cite{AEK:90} who do not
have any time delay. The coupled equations represent the interaction
between weakly nonlinear oscillators (that are near a supercritical Hopf
bifurcation) and whose coupling strength is comparable to the
attraction of the limit cycles. It is important then to retain
both the phase and amplitude response of the oscillators. Writing 
$Z_{j} = r_{j} e^{i\theta_{j}}$, Eqs.(\ref{EQN:Z1dot},\ref{EQN:Z2dot})
can also be expressed in polar form as,

\bml
\bea
\label{EQN:r1dot}
\dot{r}_{1} & = & r_{1}(1 - K -r_{1}^{2}) + K r_{2}(t-\tau)
\cos[\theta_{2}(t-\tau)-\theta_{1}], \\
\label{EQN:r2dot}
\dot{r}_{2} & = & r_{2}(1 - K -r_{2}^{2}) + K r_{1}(t-\tau)
\cos[\theta_{1}(t-\tau)-\theta_{2}], \\
\label{EQN:theta1dot}
\dot{\theta}_{1} & = & \omega_{1} + K \frac{r_{2}(t-\tau)}{r_{1}}
\sin[\theta_{2}(t-\tau) - \theta_{1}], \\
\label{EQN:theta2dot}
\dot{\theta}_{2} & = & \omega_{2} + K \frac{r_{1}(t-\tau)}{r_{2}}
\sin[\theta_{1}(t-\tau) - \theta_{2}]. 
\eea
\eml

\noindent
For $\tau =0$, the work of Aronson {\it et al.}\cite{AEK:90}
shows that the nonlinear 
Eqs.(\ref{EQN:r1dot}-\ref{EQN:theta2dot}) 
have a variety of stationary and non-stationary
solutions which depend on the strength of the 
coupling parameter $K$ and the frequency mismatch between 
the oscillators $\Delta = \mid \omega_{1} - \omega_{2}\mid$. For extremely 
weak coupling ($K \rightarrow 0$) and large $\Delta$, the 
oscillators behave independently and the long term behavior 
is a nonstationary incoherent state in which the relative 
phase of the two oscillators moves through all phases. 
Such a state is also called a {\it phase drift} state.  
With increasing coupling strength two  important classes 
of stationary solutions are possible. One of them is 
{\it amplitude death} in which the oscillators pull each other 
off their limit cycles  and collapse into the origin 
($r_{1} = r_{2} =0$) as $t \rightarrow \infty$. The other 
collective state is called {\it frequency locking} or 
{\it mutual entrainment} in which the two oscillators 
synchronize to a common frequency and the time asymptotic 
state is one of coherent or collective oscillation.
The distribution of these solutions can be  neatly represented in
a phase diagram (bifurcation diagram) in the $\Delta - K$ space.
Fig. 1, reproduced from the work of Aronson {\it et al.}\cite{AEK:90}
summarizes the above discussion. Region I represents the amplitude death
region, region III is the phase drift regime and region II marks
the phase locked state. We now analyze 
Eqs.(\ref{EQN:r1dot}-\ref{EQN:theta2dot}) for finite values of $\tau$
and examine the effect of $\tau$ on the 
conditions for the onset of these states and  changes if any
in the basic properties of these states. In the
following subsections we discuss the results for the
{\it amplitude death} and {\it phase locked} solutions.

\subsection{Amplitude Death}
\label{SUBSEC:n2death}
As is clear from Eqs.(\ref{EQN:Z1dot},\ref{EQN:Z2dot}), 
the origin $Z_{j} ~=~ 0 ~(j=1,2)$ is always a fixed point of the
system. The question to consider is whether this is a stable
fixed point in which case all amplitudes of the oscillators would
collapse into the origin as $t \rightarrow \infty$. In the absence of
time delay this state occurs in the region $K > 1$ for $\Delta > 2$.
To determine the onset
conditions of this state in the presence of time delay ($\tau \neq 0$)
we linearize Eqs.(\ref{EQN:Z1dot},\ref{EQN:Z2dot}) around $Z_{j} =0$
to obtain the characteristic equation,

\be
\label{eqn:char}
 \det ( A - \lambda I ) = 0
\ee
where $A$, the linearized matrix of
the Eqs.(\ref{EQN:Z1dot}) and (\ref{EQN:Z2dot}), is given by
\be
\label{EQN:matrixA}
A = \left[\matrix{a + i \omega_1 & K e^{-\lambda \tau} \cr
K e^{-\lambda \tau} &  a + i \omega_2 \cr} \right],
\ee
$I$ is the identity matrix, $ a = 1-K$ and the perturbations
are assumed to have a time dependence proportional to $e^{\lambda t}$.
Eqn.(\ref{eqn:char}) can be written in the form 
\bml
\be
\label{EQN:equationL}
(a - \lambda + i\omega_{1} ) ( a - \lambda + i\omega_{2} )
-K^{2} e^{-2\lambda \tau} = 0.
\ee
or
\be
\label{EQN:equationL1}
\lambda^2 - 2 (a+{\em i} \bar{\omega}) \lambda + (b_1+{\em i} b_2) + 
c e^{-2\lambda\tau} = 0
\ee
\eml
where 
$b_1 = a^2 - \bar{\omega}^2 + \Delta^2/4$, 
$b_2 = 2 a \bar{\omega}$, $\Delta = \mid\omega_1 - \omega_2\mid$, 
$\bar{\omega} = (\omega_1+\omega_2)/2$ and 
$c = - K^2$. 
This is a transcendental equation having an infinite number of
roots and we wish to study the movement of the eigenvalues in
the parametric plane of $(K, \Delta)$ and $(K, \tau)$. Setting
$\lambda = \alpha + {\em i}\beta$, where $\alpha$ and $\beta$
are real, the amplitude death region corresponds to the region in which 
$\alpha < 0$. The marginal stability curves or the critical curves 
are thus obtained by requiring that $\alpha = 0$, 
i.e. $\lambda = {\em i} \beta$. Substituting in (\ref{EQN:equationL}),
the equations defining the critical curves are thus given by,
\bml
\begin{eqnarray}
\label{EQN:forbeta1}
(\beta-\bar\omega)^{2} - \Delta^{2}/4 - a^{2} + K^{2}\cos(2\beta\tau) & = & 0, \\ 
\label{EQN:forbeta2}
2 a (\beta-\bar\omega) - K^{2}\sin(2\beta\tau) & = & 0. 
\end{eqnarray}
\eml

We first briefly describe the 
case of $\tau = 0$, 
in order to appreciate the changes brought about by finite time delay.  
Setting $\tau = 0$ in (\ref{EQN:forbeta2})
we obtain the conditions $K = 1$ and $\beta = \bar{\omega}$.
Substituting for $\beta$ in (\ref{EQN:forbeta1}) we obtain 
$K = \gamma(\Delta)= \frac{1}{2}(1+\frac{\Delta^2}{4})$. So 
the critical curves in this case are $K=1$ and 
$K = \gamma(\Delta)$ in agreement
with the work of Aronson, {\it et al.} \cite{AEK:90} and as illustrated in
Fig. 1.
It is appropriate to distinguish between two regions in the
$(K,\Delta)$ space, namely (i) $\Delta > 2$ and (ii) $\Delta < 2$.
When $\Delta > 2$ the stable region of the origin (amplitude death 
region) is bounded by $K = 1$
and $K = \gamma(\Delta)$. The eigenvalues in this particular
case can be written down (from (\ref{EQN:equationL1})) as
$\lambda = 1 - K \pm \sqrt{K^2 - \Delta^2/4} \pm i
\bar{\omega}$.  On the boundary $K = 1$, the origin loses
stability in a Hopf bifurcation. Two pairs of eigenvalues cross
into the right half plane.  On the boundary $K = \gamma(\Delta)$
a pair of eigenvalues crosses into the right hand side of the
complex eigenvalue plane, giving rise to a single frequency,
which corresponds to the phase locked state of the system. In
the second case, when $\Delta < 2$ there is no amplitude death.
However the critical curves give a boundary on which an unstable
fixed point is born as one moves to the left of $K = \gamma(\Delta)$ 
curve. Note also that the boundaries of 
the three regions meet in a highly degenerate manner in the
single point $K=1,\Delta=2$. Another distinguishing feature of
the $\tau=0$ case is that the critical curves are independent
of the mean frequency $\bar{\omega}$. In fact one can set 
$\bar \omega = 0$ (which is equivalent to transforming to
a frame rotating at the mean frequency) and carry out the
same analysis without any loss of generality. This property 
follows from the original symmetry of the coupled equations. \\

When $\tau \neq 0$, this symmetry is lost and the critical curves
are no longer independent of $\bar\omega$ as seen from 
Eqs.(\ref{EQN:forbeta1},
\ref{EQN:forbeta2}). 
In comparing the phase diagrams of the $\tau \neq 0$ case
with that of the Aronson {\it et al.}\cite{AEK:90} diagram we 
therefore need to always 
mention the specific value of the mean frequency parameter. 
We now briefly describe our numerical procedure for solving
Eqs. (\ref{EQN:forbeta1},
\ref{EQN:forbeta2}) in order to plot graphically the critical curves in the 
$(K,\Delta)$ space.
To eliminate $\beta$ between the two equations
it is more convenient to write the equations in the following parametric
form,

\bml
\begin{eqnarray}
\label{EQN:toplot1}
F & = & (\beta-\bar{\omega}) / \sin(2\beta\tau), \\
\label{EQN:toplot2}
K & \equiv & K_{\pm} = - F \pm \sqrt{F^2 + 2 F}, \\
\label{EQN:toplot3}
\Delta^2 & = & -4 a^2 + 4 (\beta-\bar{\omega})^2 + 4 K^2 \cos(2\beta\tau).
\end{eqnarray}
\eml
where (\ref{EQN:toplot2}) corresponds to the two roots of $K$ for the 
quadratic equation (\ref{EQN:forbeta2}).
Equations (\ref{EQN:toplot1}-\ref{EQN:toplot3}) represent two
sets of curves arising due to the $\pm$ signs. Let us denote the
set of curves arising due to the "$+$" sign by $S_+ \equiv
S_+(K_+,\Delta)$ and the curves due to the "$-$" sign by $S_-
\equiv S_-(K_-,\Delta)$. The curves $S_+$ and $S_-$ are obtained
by choosing an interval of $\beta$ and correspondingly
evaluating $K$ and $\Delta$ and thus eliminating $\beta$. The
function $F$ has singularities at $\beta = \beta_n = n
\pi/(2\tau)$ where $n$ is an integer. Each interval $I =
(\beta_n,\beta_{n+1})$ provides a part of the phase curves in
$(K,\Delta)$ plane. In the $(K,\Delta)$ space, the curves $S_-$
exist between $K = 1$ and $K = 2$ and the curves $S_+$ exist
outside of this region in the intervals $1/2 < K < 1$ and $K >
2$.  And at higher values of $\tau$, $S_-$ produces curves which
could intersect. For small values of the parameter $\tau$ close
to $0$, the amplitude death region is bounded by the curve $S_-$
when $K < 1$ and $S_+$ when $K > 1$. At higher values of $\tau$
the boundary of the death region falls below $K = 1$ in which
region the boundary is specified by the curves $S_+$. The
transcendental equations (\ref{EQN:toplot1}-\ref{EQN:toplot3})
must be studied for each parametric value of $\tau$ since the
curves in ($K,\Delta$) space become more complicated as the
parameter $\tau$ is increased.  \\

We now present our results of the amplitude death region for 
$\bar{\omega} = 10$ for various values of $\tau$ as obtained by the
numerical prescription described above. From the series of
diagrams in Fig. 2 we notice that with the introduction of finite
$\tau$ the point ($K=1, \Delta =2$) no longer has a degenerate
character and the critical curves begin separating and distorting
in a continuous manner.
The amplitude death region grows in size as the value of $\tau$
is increased from $0$ and the curve $S_+$ defined by the interval of 
$\beta \in (0,\bar{\omega})$ starts bending down below the 
$\Delta = 2\sqrt{2K-1}$ curve
towards the $\Delta = 0$
axis. For a critical value of $\tau = \tau_c$ it touches the $\Delta =0$
axis and 
the region of death on
the axis lies between the two points of intersection $K_1$ and $K_2$
of $S_+$ with $\Delta = 0$. This death region has a finite width $K_2 - K_1$
for a range of values of the delay parameter $\tau$.
This phenomenon of death of identical oscillators is a novel
result purely induced by the temporal delay in the coupling of
the oscillators. 
This behavior persists for a range of $\tau$ 
after which the bifurcation curve lifts up from the $\Delta =0$ line 
and starts moving upward. \\

To quantitatively study this region of amplitude death for
identical oscillators, we take a more detailed look at 
the trajectory of the two bounding points $K_1$
and $K_2$ in the parametric space of
($K,\tau$) for $\Delta = 0$. Let this trajectory be called
$\tau_b(K)$ for which we require that all the 
eigenvalues of the original transcendental equation lie in the left
half plane of the complex eigenvalue plane when $\tau >
\tau_c$. To identify such a curve from all the permissible multiple 
critical curves in the
parametric space ($K,\tau$), it is simpler to start with the original 
equation (\ref{EQN:equationL}) and set $\omega_1 = \omega_2 =
\omega_0$ in it. The eigenvalue equation now simplifies to:
\be
\lambda = 1 ~-~ K ~+~ i \omega_0 ~\pm~ K e^{-\lambda \tau}
\label{EQN:eigeneq}
\ee
Let $\lambda = \alpha + i \beta$, where $\alpha$ and $\beta$ are real.
We assume without loss of generality that $\beta \geq 0$.
In order to obtain the critical curves we set $\alpha = 0$
and consider both the equations arising out of $"+"$ and $"-"$ signs in  
(\ref{EQN:eigeneq}). After some straightforward algebra, involving 
the choice of the correct signs in the inversion of the
cosine function, we obtain the following two sets of critical curves.
\bml
\bea
\label{EQN:taukset1}
\tau_1 & \equiv & \tau_{1}(n,K) = \frac{n \pi + \cos^{-1}(1-1/K)}{\omega - \sqrt{2K-1}},\\
\label{EQN:taukset2}
\tau_2 & \equiv & \tau_{2}(n,K) = \frac{(n+1) \pi - \cos^{-1}(1-1/K)}{\omega + \sqrt{2K-1}}.
\eea
\eml
where $n = 0, 1, ..., \infty $. We further need to know the 
nature of the transition
of pairs of eigenvalues as it crosses these curves. For this it is necessary to
evaluate $d\alpha / d\tau$ on each of these curves and examine its
sign. Setting $\lambda = \alpha + i \beta$ in equation (\ref{EQN:eigeneq}) and 
differentiating with respect to $\tau$, it is straightforward to get, 
\be
\frac{d\alpha}{d\tau}\mid_{\alpha = 0} = c_1 ~\beta (\beta - \omega_0) 
= \cases{c_1 ~\sqrt{2 K -1} ~(\omega_0 + \sqrt{2 K - 1}), & if $\beta = \beta_{+}$\cr
- ~c_1 ~\sqrt{2 K -1} ~(\omega_0 - \sqrt{2 K - 1}), & if $\beta = \beta_{-}$\cr}
\ee
where $c_1 = [(1 \pm K \tau)^2 + (K \tau \sin(\beta\tau))^2]^{-1}$, which is a real 
positive constant and 
$\beta \equiv \beta_\pm = \omega_0 \pm \sqrt{2 K -1}$.
From the above equation it 
is easily seen that  
\be
\label{COND:roots}
\frac{d\alpha}{d\tau}\mid_{\alpha = 0} 
\cases{ < 0 & on $\tau_{1}$ if $K < f(\omega_0)$ \cr
        > 0 & on $\tau_{1}$ if $K > f(\omega_0)$ \cr
        > 0 & on $\tau_{2}$ \cr}
\ee
where $f(\omega_0) = (1+\omega_0^2)/2$. Thus on a 
$\tau_{1}$ curve, a pair of eigenvalues transits to the left half plane 
provided the coupling strength is smaller than $f(\omega_0)$ 
and to the right side if the coupling strength is greater than
$f(\omega_0)$. On a $\tau_{2}$ branch of the critical
curves however a pair of eigenvalues always crosses into the right half plane
of the complex plane. Thus for a finite region of amplitude death to exist
in the $K - \tau$ plane it needs to be bounded by appropriate branches of
the $\tau_{1}$ and $\tau_{2}$ curves and condition $K < f(\omega_0)$ should
hold. For $K > f(\omega_0)$, there would be no amplitude death region at all. \\

In Fig. 3, we illustrate the above arguments more graphically
by plotting the critical curves $\tau_{1}(n,K)$ (solid lines)
and $\tau_{2}(n,K)$ (dashed lines) for the first few values of $n$
and for $\bar{\omega} =10$ in the ($K-\tau$) space. 
It is possible to identify $\tau_{1}(0,K)$ as $\tau_b(K)$.
This curve forms the first boundary of the amplitude
death region of identical oscillators in ($K,\tau$) space. The stability 
of the origin can be lost when a pair of eigenvalues makes a transition 
to the right half plane, which in the present case will occur on $\tau_2(0,K)$
(as can be verified from (\ref{COND:roots})). So the region enclosed by
the intersection of $\tau_{1}(0,K)$ and $\tau_{2}(0,K)$ forms a region
of amplitude death in the $K - \tau$ space which we label as a {\it death
island}. Physically such an island represents a region in
phase space where for a given value of
$\omega$ and at a fixed $K$ we move (by varying $\tau$) from an
unstable region (corresponding to phase locked states) into a
stable region as we cross the left boundary of the island to
emerge again into a unstable region as we cross the right
boundary of the island. Is it possible to have more than one
island for a given value of $\omega_{0}$? To answer this question we see from 
Fig. 3, that the next curve along the
$\tau$ axis is $\tau_{2}(1,K)$ on which another pair of eigenvalues
will make a transition to the right half plane. Thus when one gets to
$\tau_{1}(1,K)$, which is the next in the sequence and on which a pair
of eigenvalues crosses to the left half plane, there is still a pair
left in the right half plane. Thus the region between $\tau_{1}(1,K)$
and $\tau_{2}(2,K)$ does not constitute a {\it death island}. In general
the frequency of $\tau_{2}$ curves is higher than $\tau_{1}$, i.e.,
\be
\label{EQN:N2Dorder}
\delta\tau_{2}(n,K) < \delta\tau_{1}(n,K)
\ee
where $\delta\tau_1(n,K) = \tau_1(n+1,K) - \tau_1(n,K)$ and 
$\delta\tau_2(n,K) = \tau_2(n+1,K) - \tau_2(n,K)$.
Thus the {\it death island} region is usually singly connected
and there are no higher order islands. However the ordering of
the curves depends on the magnitude of $\omega_{0}$. In the present case
we do not see any multiple islands in the range 
$4.812 \leq \omega \leq 14.438$. For $\omega_{0} > 14.438$ we
do see the appearance of higher order islands as shown in Fig. 4
for $\omega_{0} = 30$.
The size of the primary death island is found to be a function of $\omega_{0}$
and as we shall see in the next section it also depends on $N$
the number of coupled oscillators. Fig. 5, displays the island sizes
for different values of $\omega_{0}$. The size of the island
decreases with decreasing frequency and vanishes below a certain
threshold.

	All the above features of the amplitude death phenomenon have also been
confirmed by a direct numerical solution of the coupled oscillator equations
and excellent agreement with the analytic results have been found.

\subsection{Frequency Locking}
\label{SUBSEC:n2flock}
The frequency (or phase) locked solutions of the system
(\ref{EQN:r1dot}-\ref{EQN:theta2dot}) are characterized by 
the property that the relative phase of the two oscillators is 
a constant.  The phase locked state can be described by the ansatz 
$\theta_{1,2}(t) = \Omega t \pm \alpha/2$ where $\alpha$, the phase
difference between the two oscillators, and $\Omega$, the common
frequency of the two oscillators are real constants.
Substitution of 
this ansatz in Eqs.(\ref{EQN:theta1dot}) and (\ref{EQN:theta2dot}) 
further shows that the amplitudes of the limit cycles 
remain constant in this case. Thus the phase locked solutions
can be described by the representation, 
$(r_1(t),r_2(t),\theta_1(t),\theta_2(t)) = (R_1,~R_2, ~\Omega ~t - \alpha/2,
\Omega ~t + \alpha/2)$ where $R_{1,2}$ are constants.
Substituting such a form in (\ref{EQN:r1dot}-\ref{EQN:theta2dot}), 
we obtain the following set of four equations from which the
values of $R_1, R_2, \Omega$ and $\alpha$ can be evaluated.
\bml
\bea
\label{EQN:PL_R1}
(1 - K - R_1^2) R_1 + K R_2 \cos(\alpha - \Omega \tau)  = 0,\\
\label{EQN:PL_R2}
(1 - K - R_2^2) R_2 + K R_1 \cos(\alpha + \Omega \tau)  = 0,\\
\label{EQN:PL_T1}
\Omega = \omega_1 + K \frac{R_2}{R_1} \sin(\alpha - \Omega \tau),\\
\label{EQN:PL_T2}
\Omega = \omega_2 - K \frac{R_1}{R_2} \sin(\alpha + \Omega \tau).
\eea
\eml
These equations can also be rearranged in the following form
which is slightly more convenient for numerical solutions.
\bml
\begin{eqnarray}
\label{EQN:PL_1}
R_{1}^{2} & = & 1 - K + K f_{1}\cos(\alpha - \Omega \tau), \\
\label{EQN:PL_2}
R_{2}^{2} & = & 1 - K + K f_{2}\cos(\alpha + \Omega \tau), \\
\label{EQN:PL_3}
\alpha & = & \sin^{-1}\sqrt{\sin^{2}(\Omega \tau) - 
(\Omega - \omega_{1}) (\Omega - \omega_{2})/K^{2}}, \\
\label{EQN:PL_4}
R_{2}^{2} & = & f_{1}^{2}R_{1}^{2},
\end{eqnarray}
\eml
where 
$f_{1} = (\Omega - \omega_{1})/(K \sin(\alpha - \Omega \tau))$ and
$f_{2} = - ~(\Omega - \omega_{2})/(K \sin(\alpha + \Omega \tau))$.
This is a set of transcendental equations whose solutions 
describe the phase locked equilibria. We again first briefly discuss 
the $\tau = 0$ case. Putting $\tau = 0$ in 
Eqs. (\ref{EQN:PL_1}) and (\ref{EQN:PL_2}), we see that there
are two possible equilibrium solutions which are given by 
(i) $R_1^2 = R_2^2 = \rho^2$ which is called the symmetric equilibrium,
and (ii) $R_1^2+R_2^2 = 1-K$, the asymmetric case. Both these
equilibria have been studied in detail by Aronson, {\it et al.}\cite{AEK:90}. 
The symmetric phase locked equilibria are given by 
$\rho_\pm^2 = 1 - K \pm \sqrt{K^2-\Delta^2/4}$,
$\Omega = (\omega_1+\omega_2)/2$ and $\alpha_\pm$, where 
$\alpha_{+} = \sin^{-1}(\Delta/(2K))$, 
and $\alpha_{-} = \pi - \sin^{-1}(\Delta/(2K)$.
Of these two symmetric equilibria, the set given by  
$(\rho_+^2, \Omega, \alpha_+)$ is found to be stable, whereas
the solution $(\rho_-^2, \Omega, \alpha_-)$  is unstable.
The asymmetric phase locked solutions turn out to be unstable. Thus
for $\tau =0$ the only stable equilibrium is the one where
the two oscillators are synchronized to the mean frequency
and have identical amplitudes. Note that the amplitudes are
lowered from the unity value of the uncoupled case
by the amount $K - \sqrt{K^2-\Delta^2/4}$. \\

With finite time delay ($\tau \ne 0$) there is a richer
fare of equilibria. This is evident from the full set of transcendental
equations (\ref{EQN:PL_1}-\ref{EQN:PL_4}) which permit a large number of solutions. 
Although it
is difficult to obtain analytic forms for these solutions, it is easy
to track these solutions numerically. We have carried out such an analysis
and our results are displayed in Fig. 6 for $\tau = 0.4084$. 
The number of these coherent states increases
as a function of $K$ and $\tau$. We have also studied the stability
of these states by carrying out a linear perturbation analysis of
Eqs. (\ref{EQN:r1dot}-\ref{EQN:theta2dot}) around the phase locked 
solutions. The algebraic details of this analysis are given
in Appendix. Numerical solution of the stability conditions
shows that all these higher frequency states are stable 
(except for small bands of unstable regions which are indicated
by dashed curves in Fig. 6). Thus these higher frequency states are 
genuine collective states of the system that are physically accessible.
It should be mentioned that similar collective states were also
observed in the delay coupled {\it phase only} model studied by
Schuster {\it et al.}\cite{SW:89}. One difference between those results and
our amplitude inclusive model is that the frequencies
of our model are slightly lower due to amplitude effects.
When the phase locked amplitudes of the two oscillators 
are equal or if their ratio is close to unity, the effect of the 
amplitude on the magnitude of the phase locked frequencies ceases.  \\

We end this analysis of the $N=2$ model with a short description
of the typical phase diagram for finite time delays. Such a
diagram is shown in Fig. 7 for $\tau= 0.4084$ which reveals
a much richer structure in comparison to the
Aronson {\it et al}\cite{AEK:90} diagram of Fig. 1.
Note that one no longer has the clean separation of the Bar-Eli
region, the phase locked region and the phase drift region into
three disjoint regions that converge at a single degenerate
point. Instead the phase locked region now always surrounds the Bar-Eli
region and the single degenerate point is replaced by a
series of $X$ points resulting from the braided structure of the
phase locked region in the vertical direction. The dotted curve
(obtained numerically) marks the separation of the phase locked
and the incoherent regions. This curve also represents the birth
of two fixed points of the system, one stable and the other
unstable. These branches can be seen in Fig. 8a where the phase
locked amplitudes are plotted at $\Delta = 2$. The dashed curves
are the unstable branches and the solid curves stable branches.
At large values of $K$ other bifurcation curves appear in the
phase locked region indicating the appearance of higher
frequency states \cite{SW:89}. Figs. 8b and 8c show respectively the 
phase locked solutions for $\Delta = 7$ as $K$ is varied and
for $K = 1$ as $\Delta$ is increased. It should be noted that the
basic nature of the transitions, namely a supercritical Hopf bifurcation,
is preserved in the presence of time delay. Fig. 8b, a typical
example, illustrates this clearly.

\section{Assembly of N Delay Coupled Oscillators}
\label{SEC:ngen}
\subsection{Model Equations}
\label{SUBSEC:ngenmodel}

In this section we study the interaction of a
large number of limit cycle oscillators that are globally coupled
with a linear time delayed coupling. To describe such a system,
we introduce the following set of model
equations,
\bea
\label{EQN:ZdotN}
\dot{Z}_j(t)=(1+i\omega_j-\mid Z_j(t)\mid^2)Z_j(t)
+ & ~\frac{K^\prime}{N}\sum\limits_{k = 1}^{N}[Z_k(t-\tau)-Z_j(t)] \nonumber \\
- & ~\frac{K^\prime}{N}[Z_j(t-\tau)-Z_j(t)]
\eea
where $j = 1, ..., N$,  $K^\prime = 2 K$ is the coupling strength and
$\tau$ is the delay time. The last term on the R.H.S. has been
introduced to remove the self-coupling term. The model is a
straightforward generalization of the mean field model studied
extensively by Ermentrout\cite{Erm:90}, Mirollo and Strogatz
\cite{MS:90a}, and others\cite{Aiz:76} and reduces exactly to
their model for $\tau=0$. Mirollo and Strogatz\cite{MS:90a} have
provided rigorous analytical and numerical conditions for
amplitude death in such a mean field model system.  Their
conclusions, in general, are similar to the case of $N=2$,
namely, that one needs a sufficiently large variance in
frequencies for death to occur and $K$ has to be sufficiently
large. In a short while we will discuss our findings for the time
delayed model in the light of their results. \\

As is customary in mean field models we can also define
a centroid or `order parameter' of the form,
\be
\bar{Z} = R ~e^{i ~\phi} ~=~ \frac{1}{N} ~\sum_{j = 1}^{N} ~Z_j(t)
\ee
where $R$ and $\phi$ denote the amplitude and phase of the centroid.
The order parameter is a useful quantity in the large $N$ model
since its behavior provides  qualitative and quantitative clues about
the collective and nonstationary states of the system, e.g. $R=0$
(in the large time limit) is indicative of an incoherent state
whereas $R=1$ marks a  `phase locked' state. As has been noted by
Matthews and Strogatz\cite{MS:90,MMS:91} the time behavior of $R$ can
also characterize chaotic states and other nonstationary
states like large amplitude Hopf oscillations. We will also
examine the behavior of this parameter in certain cases to
track the effects of the time delay parameter on the collective
dynamics of the large $N$ system.  In terms of the order
parameter defined above the model equations can be written as
follows.
\be
\label{EQN:ZdotNbar}
\dot{Z}_j(t)=(1 - K^\prime d + i\omega_j-\mid Z_j(t)\mid^2)Z_j(t)
+ ~K^\prime \bar{Z}(t-\tau)
- ~\frac{K^\prime}{N} Z_j(t-\tau) 
\ee
where  $d = 1 - 1/N$. 
To study the stability of the origin, we once again 
carry out a linear perturbation analysis of (\ref{EQN:ZdotN})
with the perturbation varying as $e^{\lambda t}$. The 
linearized matrix of (\ref{EQN:ZdotN}) is given by
\be
\label{EQN:N2DmatrixB}
B = \left[\matrix{
l_1& 
f & 
\cdots & 
f \cr
f & 
l_2& 
\cdots & 
f \cr
\vdots & \vdots & \ddots & \vdots \cr
f &
f &
\cdots & 
l_N\cr
} \right],
\ee
where $l_n = 1 - K^\prime d + i \omega_n$,
$f = \frac{K^\prime}{N} ~e^{-\lambda \tau}$. That is 
$$
B_{ij} = \cases{1 - K^\prime d + i \omega_j, & if $i ~=~ j$ \cr
                \frac{K^\prime}{N} e^{-\lambda \tau}, & if $i ~\ne~ j$ \cr }
$$
It is more convenient to cast the
eigenvalue problem in terms of another matrix $C$ where we
define $C = B + (K^\prime d - 1) I$ (with $I$ being the identity matrix).
If $\mu$ is the eigenvalue of $C$ then it is 
related to $\lambda$ by the relation, $\mu = \lambda +
(K^\prime d -1)$.The matrix
$C$ is given by 
\be
C_{mn} = \cases{i \omega_m, & if $m = n$, \cr
                f, & if $m \ne n$. \cr}
\ee
The eigenvalues $\mu$ 
are obtained by solving $\det(C - \mu I) = 0$, i.e.,
\be
\label{EQN:matrixCMU}
\det \left[\matrix{i \omega_1 - \mu & f & \cdots & f\cr
f & i \omega_2 - \mu & \cdots & f \cr
\vdots & \vdots & \ddots & \vdots \cr
f & f & \cdots & i \omega_N - \mu \cr}\right] = 0,
\ee
This eigenvalue equation can be compactly expressed as a product
of two factors,
\be
\label{EQN:spect}
\left [ \prod_{k = 1}^{N} (i \omega_k - \mu - f)\right ] \left[ 1 + 
f \sum_{j = 1}^{N}\frac{1}{i \omega_j - \mu - f} \right] = 0
\ee
As pointed out by Matthews and Strogatz\cite{MS:90,MMS:91}, the first factor 
represents the continuous spectrum 
of the system whereas the second factor gives us the discrete spectrum. 
In general it is not possible to solve the characteristic equation
(\ref{EQN:spect}) analytically and for large $N$ the numerical
tracking of all the eigenvalues is also an arduous task. There
are two interesting limits however, in which the analysis gets
considerably simplified. If the $N$ oscillators have identical
frequencies then it is possible to obtain exact algebraic
relations for the critical curves marking the amplitude death
region.  This is of considerable interest to us in view of the
novel result of the $N=2$ model which showed amplitude death for
$\Delta =0$.  The interesting question to ask is whether such a
phenomenon exists for the arbitrary $N$ case. We will study this
question in the next subsection. Another interesting limit is
the $N \rightarrow \infty$ limit, often called the thermodynamic
limit. It is once again possible to obtain some exact analytic
results in this limit regarding the behavior of the critical
curves in the bifurcation diagram. We will carry out this
analysis in the subsection \ref{SUBSEC:ngeninfdeath}.  As shown by 
Mirollo and Strogatz\cite{MS:90a} (for the $\tau=0$ case) the infinite system 
results are of practical interest since they provide a fairly accurate
picture of amplitude death for the large but finite system. A
similar conclusion holds for the time delay case as well, as we
find out by comparing the analytic results of the thermodynamic limit
with numerical solutions for a large number of oscillators.
In the final subsection we look at the time behavior of the order
parameter in the various collective regimes and discuss its
dependence on the time delay factor. We also briefly
discuss the region  of nonstationary solutions
(chaos, Hopf oscillations etc.) that was first pointed out by 
Matthews and Strogatz\cite{MS:90,MMS:91} for the $\tau=0$ case.

\subsection{Amplitude death of identical oscillators for arbitrary N}
\label{SUBSEC:Ngendeath}

In this section we treat the case of identical
oscillators suffering amplitude death for an arbitrary number 
of globally coupled
oscillators, $N$. 
For a set of identical oscillators the frequency
distribution of the system is a delta function,
\be
\label{EQN:N2Dg}
g(\omega) = \delta(\omega-\omega_0)
\ee
where $\omega_{0}$ is the natural frequency of each oscillator.
With this assumption the matrix $B$ becomes a circulant matrix 
whose eigenvalues can be expressed in terms of the $N^{th}$ roots 
of unity. Since in this particular case only two kinds of 
coefficients appear in the matrix $B$, the eigenvalues become 
much simpler. Explicitly they are given by
\bea
\label{EQN:N2Dev}
\lambda = \{ 1 - K^\prime d +i \omega_0 + K^\prime d e^{-\lambda \tau},
\;\; 1 - K^\prime d +i \omega_0 - \frac{K^\prime}{N} e^{-\lambda \tau} \},
\eea
in which the second eigenvalue has a degeneracy of $N-1$.
Considering both the eigenvalue equations and following a similar procedure
as described for the $N=2$ case, we obtain the following set of
critical curves:
\bml
\bea
\label{EQN:A}
\tau_a(n,K) & = & \frac{2 n \pi + \cos^{-1}\left[1-\frac{1}{K^\prime d}\right]}{\omega_0 - \sqrt{2 K^\prime d -1}},\\
\label{EQN:B}
\tau_b(n,K) & = & \frac{2 (n+1) \pi - \cos^{-1}\left[1-\frac{1}{K^\prime d}\right]}{\omega_0 + \sqrt{2 K^\prime d -1}},\\
\label{EQN:C}
\tau_c(n,K) & = & \frac{2 (n+1) \pi - \cos^{-1}\left[\frac{1-K^\prime d}{K^\prime(1-d)}\right]}{\omega_0 - \sqrt{[K^\prime (1-d)]^2 - (K^\prime d -1)^2}},\\
\label{EQN:D}
\tau_d(n,K) & = & \frac{2 n \pi + \cos^{-1}\left[\frac{1-K^\prime d}{K^\prime (1-d)}\right]}{\omega_0 + \sqrt{[K^\prime (1-d)]^2 - (K^\prime d -1)^2}}.
\eea
\eml
Note that $N$ enters as a parameter (through $d = 1 - 1/N$) in these sets of curves
and the family of curves is now four instead of the two found for the 
$N=2$ case. In fact for $N = 2$, the curves $\tau_a(n,K)$ and $\tau_c(n,K)$ combine 
to give $\tau_1(n,K)$ and $\tau_b(n,K)$ and $\tau_d(n,K)$ combine to give
$\tau_2(n,K)$ of the previous section. In Fig. 9 we display some typical
death island regions for various values of $N$ as obtained from the
critical curves (\ref{EQN:A})--(\ref{EQN:D}) with $n=0$. The sizes of the islands
are seen to vary as a function of $N$ and approach a saturated size as 
$N \rightarrow \infty$. The existence of these regions independently confirmed
by direct numerical solution of the coupled oscillator equations, demonstrates
that the amplitude death phenomenon for identical oscillators happens
even in the case of an arbitrarily large number of oscillators. They also
display multiple connectedness of the death region for higher values of $\omega_0$
as was seen for the $N = 2$ case.

\subsection{Oscillator death in the thermodynamic limit} 
\label{SUBSEC:ngeninfdeath}
In the thermodynamic limit ($N \rightarrow \infty, ~d \rightarrow 0$) the summation
in the discrete spectrum of Eqn. (\ref{EQN:spect}) can be replaced by an integral and
replacing $\mu$ by its definition, the discrete eigenvalue equation
can be written as,
\be
\label{EQN:NDINFchar1}
h(\lambda) = \int_{-\infty}^{\infty} \frac{g(\omega)}
{\lambda + K^\prime - 1 - i \omega} d\omega
= \frac{1}{K^\prime} e^{\lambda \tau}
\ee
where $g(\omega)$ denotes the distribution of the frequencies.
The continuous spectrum is 
given by
\be
\label{EQN:NDINFchar2}
\lambda = 1 - K^\prime 
+ i \omega ~~ where ~~~\omega  \in ~~g(\omega)
\ee
From (\ref{EQN:NDINFchar2}) one can see that 
one of the critical curves is $K^\prime = 1$. And the amplitude death 
region should lie to the right of this region in order for the 
eigenvalues to have negative real parts. To determine the critical 
curves from the discrete spectrum, we substitute $\alpha + i \beta$
for $\lambda$ in (\ref{EQN:NDINFchar1}), rationalize the denominator 
and equate the real and imaginary parts to zero, to obtain at $\alpha = 0$,
\bml
\bea
\label{EQN:NDINFcharI1}
I_1 & ~~=~~ & \int_{-\infty}^{\infty}\frac{K^\prime-1}{(K^\prime-1)^2 
+ (\beta-\omega)^2} 
~~g(\omega) ~~d\omega 
= \frac{1}{K^\prime} ~~\cos(\beta\tau),\\
\label{EQN:NDINFcharI2}
I_2 & ~~=~~ & \int_{-\infty}^{\infty}
\frac{\omega-\beta}{(K^\prime-1)^2 + (\beta-\omega)^2} 
~~g(\omega) ~~d\omega
= \frac{1}{K^\prime} ~~\sin(\beta\tau).
\eea
\eml
These two relations define the other critical curve and for
a given $g(\omega)$ they provide a complete analytic description.
As an example, let us choose a simple uniform distribution given by,
\be
g(\omega) ~=~\cases{\frac{1}{2 \sigma}, & if $\omega ~\in ~~[-\sigma, \sigma]$ \cr
                 0, & otherwise \cr }
\ee
The integrals $I_1$ and $I_2$ can now be explicitly carried out
and we get,
\bml
\bea
\label{EQN:NDINFtan}
\tan^{-1}\left(\frac{\sigma+\beta}{K^\prime -1}\right) 
+ \tan^{-1}\left(\frac{\sigma-\beta}{K^\prime -1}\right)
& = & \tan^{-1}\left(\frac{2 \sigma ~(K^\prime-1)}{(K^\prime-1)^2 
~-~ \sigma^2 + \beta^2}\right) \nonumber \\
& ~=~ & \frac{2 ~\sigma}{K^\prime} ~ \cos(\beta \tau), 
\eea
and
\be
\label{EQN:NDINFlog}
\log\left(\frac{(K^\prime-1)^2~+~(\sigma-\beta)^2}{(K^\prime-1)^2
~+~ (\sigma+\beta)^2}\right)
= \frac{4 ~\sigma}{K^\prime} ~\sin(\beta \tau)
\ee
\eml
Thus in the limit of an infinite number 
of oscillators the above set of transcendental equations,
parametrized by the time delay, provide a complete description of
the critical curve for amplitude death (along with $K^\prime = 1$).
It is also easy to see that for $\tau = 0$,
$\beta = 0$ is the solution
one obtains from (\ref{EQN:NDINFlog}) and upon substitution of
this in (\ref{EQN:NDINFtan}) one obtains %
\be
\tan^{-1}\left(\frac{\sigma}{K^\prime - 1}\right) ~=~ \frac{\sigma}{K^\prime}
\ee
which is the equation for the critical curves in the no delay case 
as was obtained by \cite{Erm:90,MS:90a}.  In Fig. 10  we plot the 
region of amplitude death for a typical value of $\tau$.
We also note
that the general nature of the bifurcation curves is similar
to the $N=2$ case including the occurrence of amplitude
death for $\Delta =0$. As mentioned earlier,
Mirollo and Strogatz\cite{MS:90a} also point out that the 
infinite system provides a
fairly accurate description of the large finite system. In order
to check this out for the time delayed system we have carried 
out numerical studies of a large system ($N=800$) of oscillators
for a few values of $K$ and $\Delta$ at a fixed value of $\tau$.
The values of $\Delta$ and $K$ at which the transition from the
phase locked region to the death region occurs are marked with
circles in Fig. 10. As we see these points lie nearly
on the analytic curves for the $N=\infty$ case and thus the
bifurcation diagram for the finite large system can be expected
to be similar to the infinite case.

\subsection{Behavior of the order parameter}
\label{SUBSEC:orderparameter}
It is also interesting to examine the behavior of the
order parameter in the limit of large number of oscillators
since it provides a good description of the macroscopic
behavior of the system. Our particular interest now is to
detect any signatures of the time delay effect that might
show up in the temporal behavior of the order parameter. \\

We begin by studying its characteristic behavior in 
the amplitude death region where it is known to spiral
in time to the origin. To investigate the influence of
the time delay we can measure this exponential decay rate
for various values of $\tau$. Our numerical results are
plotted in Fig. 11 where the decay exponent is plotted against
$\tau$ for two values  of $K$ and for a particular value
of $\omega$ for the case of $N=800$ oscillators. The range of
$\tau$ spans the width of a death island. The exponent is seen
to have a maximum negative value in the centre of the island and
it gets less negative as one moves towards the boundaries. Thus
time delay has a dynamical influence on the rate at which
the oscillators collapse into the origin with the maximum
rate of decay occurring in the centre of the island. \\

The death islands are surrounded by the phase locked regions in
which the order parameter assumes a non zero constant value.
The magnitude of this value is found to be a function of $\tau$.
Our numerical results show that the saturated value of the order 
parameter grows in an algebraic fashion as a function of $\tau$
as we move from the death island boundary into the phase locked
region. This is demonstrated in Fig. 12  for $K^\prime =5$. \\

Finally we examine the behavior of the order parameter in the
narrow region straddling the boundary between the phase locked and
the phase drift region where Matthews and Strogatz\cite{MS:90,MMS:91}
have found non-stationary
behavior corresponding to Hopf oscillations, large oscillations,
quasiperiodicity and chaos. We find that the order parameter
displays all these characteristic behaviors even in the presence of
the time delay. However there is an overall reduction of the
phase space area of this region compared to the no delay case.
To see this clearly we choose to consider the temporal record of the
amplitude of the order parameter. The value of the order parameter
is constant in the phase locked region, and tends to zero as $1/\sqrt{N}$
with $N$, the number of the oscillators, in the incoherent region. 
In the irregular region a variety of behaviours can be seen. 
We take a long time record of $R$ discarding the first few hundred points.
The average value of this record, $<R>$,
as shown in Fig. 13, for $N = 500$, shows a plateau in the irregular region.
This is almost unchanged when the number of oscillators is increased
from $N = 500$ to $N = 5000$. This region shrinks considerably as the
delay parameter $\tau$ is increased. 
We have also observed other qualitative 
changes in the nature of the time variation of the order parameter
which suggest subtle influences of time delay on 
the dynamical pattern of the nonstationary states. A more detailed
and quantitative study of these effects are in progress and will
be reported elsewhere. \\

\section{Summary and Discussion}
\label{SEC:summary}
We have studied the effect of time delay on the collective dynamics
of a system of coupled limit cycle oscillators that are close to
a supercritical Hopf bifurcation. The oscillators are linearly coupled 
in a global fashion and time delay is introduced in the coupling term. 
The principal effects of time delay on the stationary collective states
like amplitude death and phase locking are clearly evidenced even
in a simple two coupled oscillator system. The phase diagram shown
in Fig. 7 summarizes these results. The boundary curve for the
amplitude death region is a sensitive function of the time delay
parameter and the phase space distribution of locked states,
amplitude death and incoherent states is quite complex. The phase
locked region shows the existence of multiple locked states all of
which are stable. A surprising result is the occurrence of amplitude
death even for two identical oscillators over a range of time delay
values (Fig. 5). Such death islands in the parametric space of the coupling
strength and the time delay factor are found to persist even
for a large number of identical oscillators. Our numerical findings
are backed by analytic results of linear stability of 
the amplitude death and phase locked states for the $N=2$ and
the $N \rightarrow \infty$ cases. The large $N$ results are found to agree
very closely with the $N \rightarrow \infty$ curves. For the case of identical
oscillators we provide exact analytic curves for the island boundaries.
Our principal focus has been on the investigation of the stationary
states but we have also confirmed the existence of the nonstationary
states (e.g. Hopf oscillations, chaos) in the finite time delay
model. We hope to carry out a more detailed investigation of these
states in the future.\\ 

As we have mentioned earlier, time delay effects in the
context of the coupled oscillator model have not been looked
at before. In view of the vast number of applications of
the coupled oscillator model our results may have 
significance for some biological or physical systems. There are many 
physical examples of amplitude death in real systems.
One of the earliest that was investigated both theoretically
and experimentally is that of coupled chemical oscillator
systems e.g. coupled Belousov-Zhabotinskii reactions carried out
in coupled stirred tank reactors\cite{Bar:85,CE:89}. They can
also occur in ecological contexts where one can imagine two
sites each having the same predator-prey mechanism which causes
the number density of the species to oscillate. If the species
are capable of moving from site to site at a proper rate
(appropriate coupling strength) the two sites may become stable
(stop oscillating) and acquire constant populations. Another
important application of this concept is in pathologies of
biological oscillator networks e.g. an assembly of cardiac
pacemaker cells\cite{Win:80}.  Amplitude death signifies cessation of
rhythmicity in such a system which is otherwise normally
spontaneously rhythmic for other choices of parameters.  For the
onset of such an arrhythmia, current models based on coupled
oscillator networks need to assume a significant spread in the
natural frequencies of the constituent cells (oscillators)\cite{MS:90a}. Our
result of amplitude death for identical oscillators demonstrates
that this assumption may not be necessary if one takes into
account time delay effects arising naturally from the finite
propagation times of the signals exchanged between the cells.
Another possible application is in the area of high power
microwave sources where it is proposed to enhance the microwave
power production by phase locking a large number of sources such
as relativistic magnetrons\cite{BSWSH:89}. Time delay effects,
arising from the finite propagation time of information signals
traveling through the connecting waveguide bridges, could impose
important limitations on the connector lengths and geometries in
these schemes. Our findings could provide a guideline in this
direction. Coupled oscillators also feature in several neural network
configurations where multiple phase locked states arising
from time delay effects could play an important role. \\

Finally we would like to discuss some further extensions of
the present work and possible future directions of research in
this area. Our present results have been largely derived from
the simple model of limit cycle oscillators, which are close to
a supercritical Hopf bifurcation and which are linearly
coupled in a global fashion. It would be interesting to carry
out a similar analysis for limit cycle oscillators with local
coupling. For the particular case of death of identical
oscillators we have confirmed that the result holds for the
locally coupled model as well but a more general investigation
of other collective states remains to be done.
Likewise the introduction
of more complicated dynamics in the individual oscillators,
such as shear (amplitude dependent frequency)\cite{PM:92} and higher order 
nonlinearities can open up rich possibilities for the 
interplay of time delay and the collective dynamics of the system.

\appendix
\section{Stability of the Frequency Locked Solutions}
\label{SEC:appendix}

The phase locked solutions for the case of $ N = 2 $ are given by 
$(\psi_1,\psi_2,\psi_3,\psi_4) = 
(R_1, R_2, \Omega t + \alpha_1, \Omega t + \alpha_2)$.
To determine the stability of the phase locked states we carry 
out a linear perturbation analysis about 
these solutions. Suppose $\lambda$ represents the eigenvalue. The 
characteristic equation is then given by
\be
\label{EQN:APPENDm1}
\det\left|\matrix{
A-\lambda & C_1 e^{-\lambda\tau} & R_2 D_1 &  - R_2 D_1 e^{-\lambda\tau} \cr
C_2 e^{-\lambda\tau} & B - \lambda & -R_1 D_2 e^{-\lambda\tau} & R_1 D_2 \cr
-\frac{R_2}{R_1^2}D_1 & \frac{D_1}{R_1} e^{-\lambda\tau} & 
\frac{-R_2}{R_1}C_2 - \lambda & \frac{R_2}{R_1}C_1 e^{-\lambda\tau} \cr
\frac{D_2}{R_2}e^{-\lambda\tau} & -\frac{R_1}{R_2^2}D_2  & 
\frac{R_1}{R_2} C_2 e^{-\lambda\tau} &  - \frac{R_1}{R_2} C_1 -\lambda \cr
} \right| = 0
\ee
where $A = 1 - K - 3 R_1^2, ~~B = 1 - K - 3 R_2^2, 
~~C_1 = K \cos(-\Omega\tau+\alpha), ~~C_2 = K \cos(\Omega\tau+\alpha),
~~D_1 = K \sin(-\Omega\tau+\alpha), ~~D_2 = K \cos(-\Omega\tau-\alpha)$.
We can also write the above determinant in the form of the 
following equation:
\be
\label{stabeq}
f_{00} + f_{01} \lambda + f_{02} \lambda^2 + f_{04} \lambda^4
       + (f_{20} + f_{21} \lambda + f_{22} \lambda^2) e^{-2 \lambda \tau}
                 +  f_4  e^{-4 \lambda \tau} = 0
\ee
where the coefficient functions $f_{ij}$ are given by
\bea
f_{00} & =&  A B C_1 C_2 + D_1^2 D_2^2 - A C_2 D_2^2 R_1/R_2 - B C_1 D_1^2 R_2/R_1,\\
%
%
f_{01}  &=&  - A C_1 C_2 - B C_1 C_2 - A D_2^2 R_1^2/R_2^2 - B D_1^2 R_2^2/R_1^2 \nonumber \\ 
        & + & A B C_1 R_1/R_2 + A B C_2 R_2/R_1  + C_2 D_2^2 R_1/R_2 + C_1 D_1^2 R_2/R_1, \\
%
%
f_{02} &=& A B + C_1 C_2 + D_1^2 R_2^2/R_1^2 + D_2^2 R_1^2/R_2^2 - A C_1 R_1/R_2 - B C_1 R_1/R_2 \nonumber \\
       &-& A C_2 R_2/R_1 - B C_2 R_2/R_1, \\
%
%
f_{04}  & =&    - A - B + C_1 R_1/R_2 + C_2 R_2/R_1,\\
%
%
f_{20} &=& - A B C_1 C_2 - C_1^2 C_2^2 - C_1 C_2 D_1^2 - C_1 C_2 D_2^2 \nonumber \\
       &-& C_1^2 D_1 D_2 - C_2^2 D_1 D_2 + 2 C_1 C_2 D_1 D_2 - 2 D_1^2 D_2^2 \nonumber\\
       &-& A C_1 D_1 D_2 R_1/R_2 + A C_2 D_1 D_2 R_1/R_2 + B C_1 D_1 D_2 R_2/R_1 - B C_2 D_1 D_2 R_2/R_1 \nonumber \\
       &+& A C_1 D_2^2 R_1/R_2 + B C_2 D_1^2 R_2/R_1, \\
%
%
f_{21} &=& A C_1 C_2 + B C_1 C_2 - A D_1 D_2 - B D_1 D_2 - C_1^2 C_2 R_1/R_2 - C_1 C_2^2 R_2/R_1 \nonumber\\
       &+& C_1 D_1 D_2 R_1/R_2 + C_2 D_1 D_2 R_2/R_1 - 2 C_2 D_1 D_2 R_1/R_2 - 2 C_1 D_2^2 R_1/R_2  \nonumber \\
       &-&  2 C_2 D_1^2 R_2/R_1 - 2 C_1 D_1 D_2 R_2/R_1, \\
%
%
f_{22} & = & - 2 C_1 C_2 + 2 D_1 D_2, \\
%
%
f_4 & =&  C_1^2 C_2^2 + C_2^2 D_1^2 + C_1^2 D_2^2 + D_1^2 D_2^2 = 1.
\eea
The above equation (\ref{stabeq}) has been solved numerically to
generate the stability curves displayed in the paper.

\newpage


%
%
\vfill
\centerline{ {\Large Figures} }
\vfill
\begin{figure}
\centerline{\psfig{file=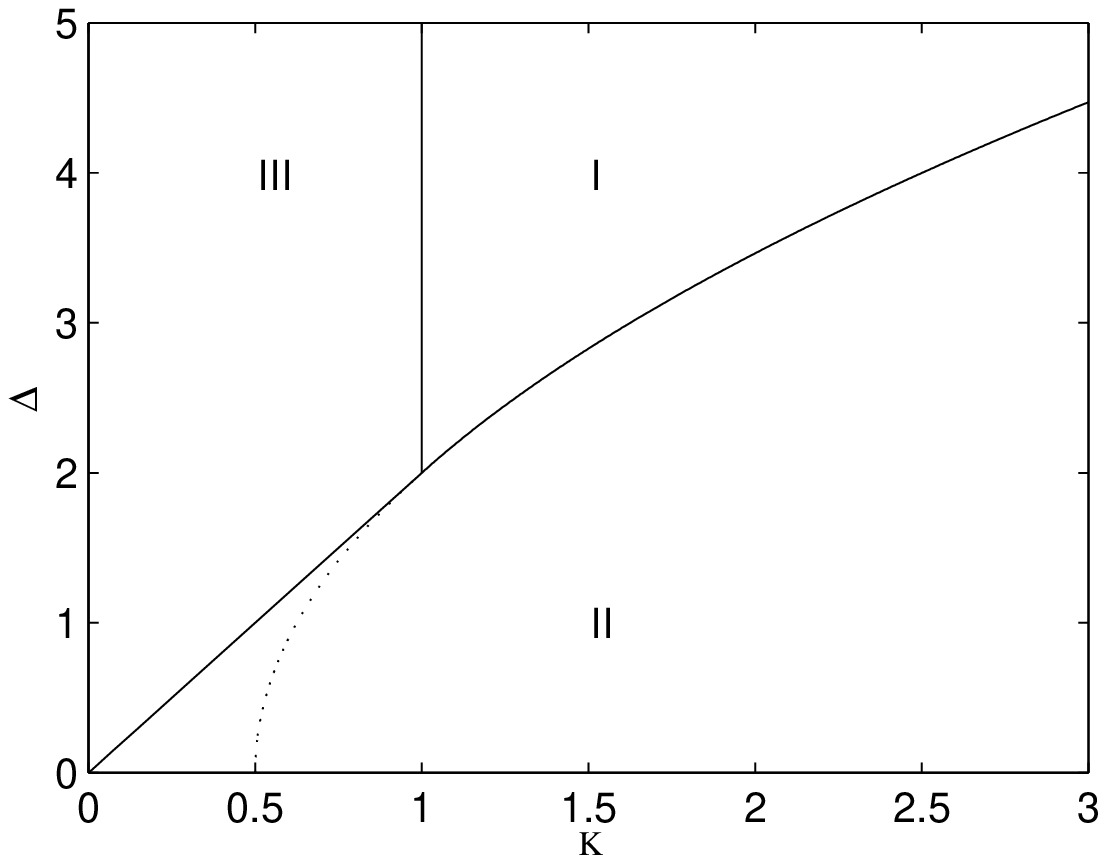,width=8cm,height=8cm}}
\caption{Bifucrcation diagram of Aronson, {\em et al.} of two coupled oscillators without
time delay in the plane of coupling strength ($K$) and the frequency mismatch ($\Delta$).
Region I, bounded by the curves $K = 1$ and $K = (1+\Delta^2/4)/2$,
represents the oscillator death region. Region II, bounded by $K = \Delta/2$ 
(when $\Delta < 2$) and $K = (1+\Delta^2/4)/2$ (when $\Delta > 2$), 
represents the frequency locked state. Region III represents the incoherent 
or drift solutions. All three regions meet at a degenerate point $(K,\Delta) = (1,2)$.}
\end{figure}

\begin{figure}
\centerline{\psfig{file=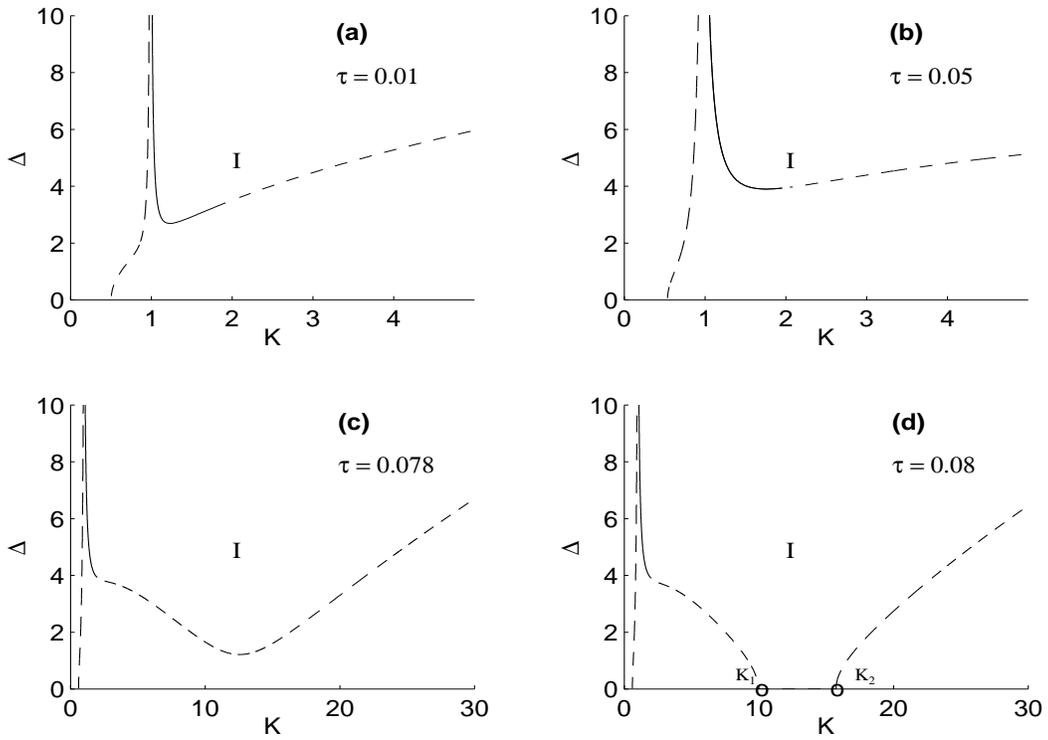,width=14cm,height=10cm}}
\caption{The effect of time delay on the boundary of the amplitude death region of 
two coupled oscillators. The critical curves are plotted 
from Eqs. (\ref{EQN:toplot1}--\ref{EQN:toplot3}) for $\bar{\omega} = 10$. Region 
I represents the amplitude death region. For small 
values of $\tau$ the denerate point $(K,\Delta) = (1,2)$ disappears and the
death region is bounded by the curves $S_{-}$ (solid lines) 
and $S_{+}$ (dashed lines) as defined in the text. As the
delay parameter $\tau$ is increased, the bounding curves get deformed 
continuously.}
\end{figure}

\begin{figure}
\centerline{\psfig{file=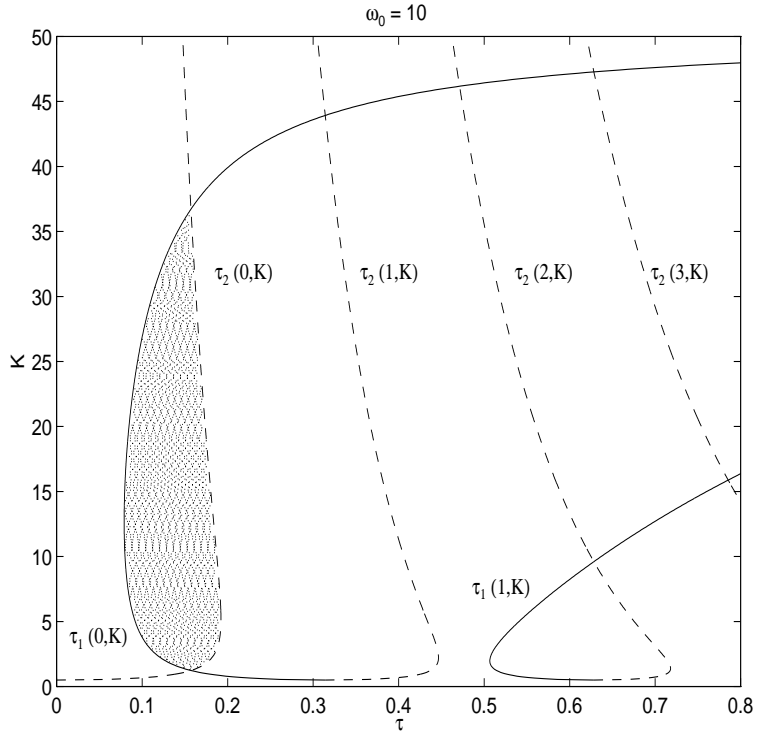,width=10cm,height=10cm}}
\caption{The {\em death island}. The amplitude death region for identical oscillators
with a common frequency of $\bar{\omega}\equiv\omega_0 = 10$ in $(K,\tau)$
space. The island boundaries are defined by $\tau_1(0,K)$ and $\tau_2(0,K)$.
No other regions of amplitude death region exist for this value of $\omega_0$.}
\end{figure}

\begin{figure}
\centerline{\psfig{file=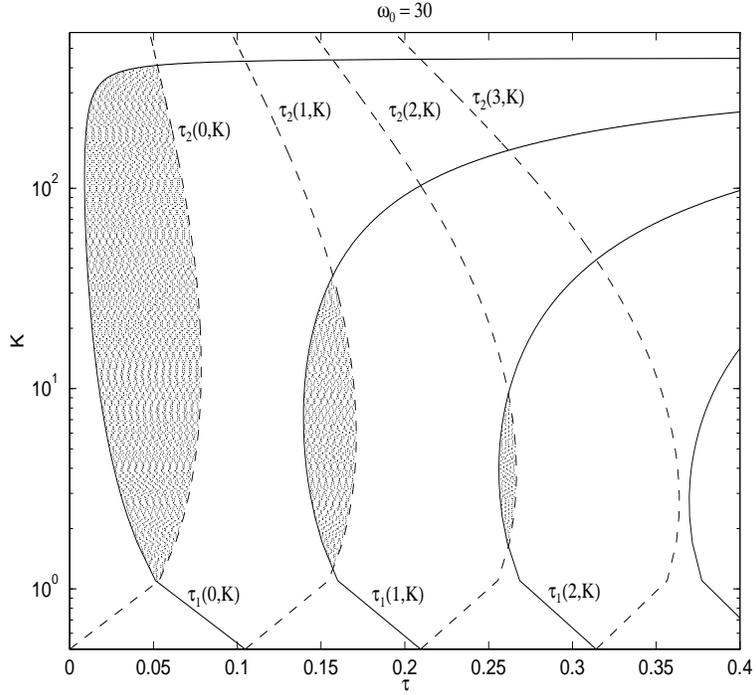,width=10cm,height=10cm}}
\caption{The existence of multiple death islands. In the $(\tau,K)$ space
the amplitude death region is multiply connected (i.e. higher order death islands
exist)  for higher values of $\omega_0$. The figure shows that for $\omega_0 = 30$,
there are three death islands which are defined by the set of curves $\tau_{1,2}(m,K)$
where $m = 0, 1, 2$.}
\end{figure}

\begin{figure}
\centerline{\psfig{file=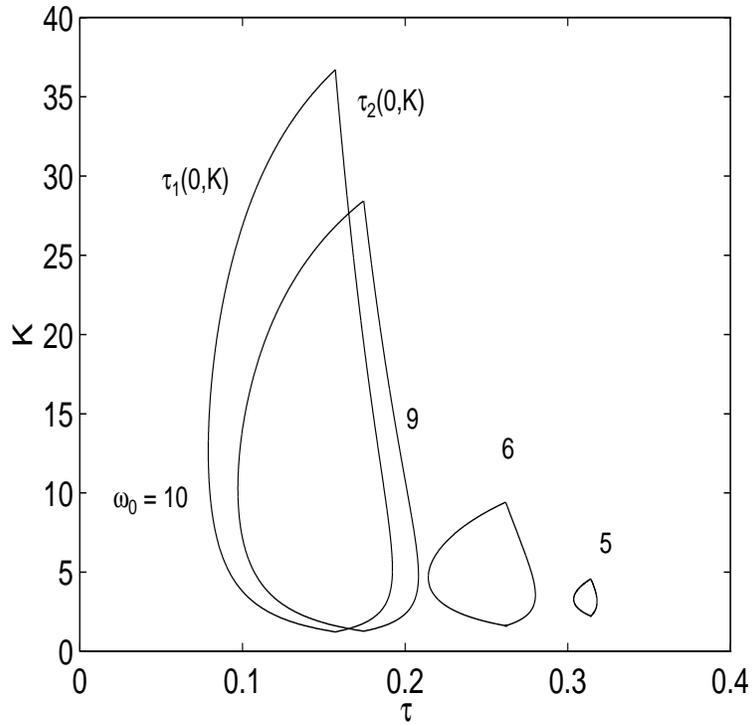,width=10cm,height=10cm}}
\caption{Dependence of the size of the death island on the common intrinsic frequency,
$\omega_0$. Below a certain threshold, which is given by the condition of intersection of
the curves $\tau_1(0,K)$ and $\tau_2(0,K)$, the amplitude death region 
disappears. This value, found numerically, is $4.812$.}
\end{figure}

\begin{figure}
\centerline{\psfig{file=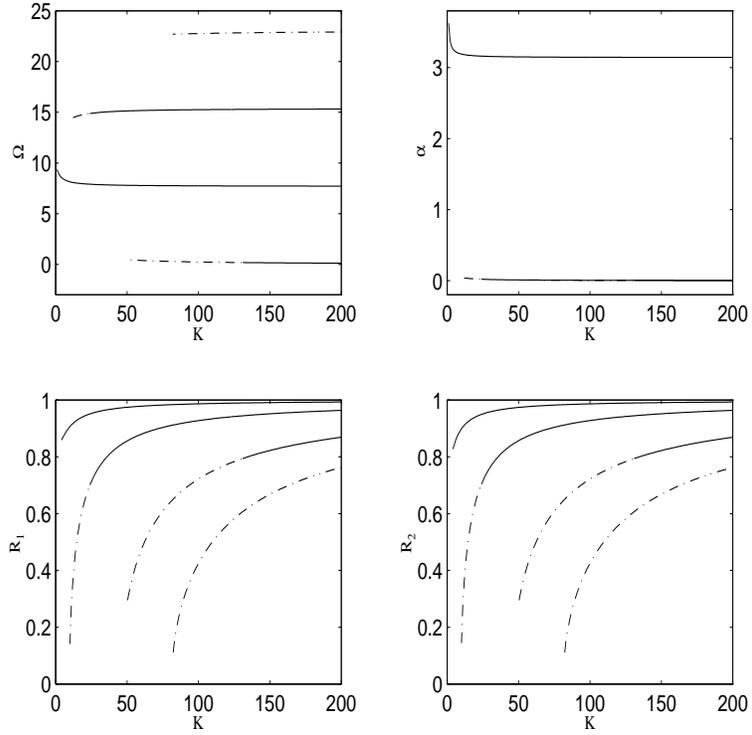,width=10cm,height=10cm}}
\caption{The phase locked solutions. 
The common frequency $\Omega$, the phase difference $\alpha$ and
the amplitudes of the individual oscillators plotted as a function of the coupling 
strength, $K$ for $\tau = 0.4084$, $\Delta = 1$,  and $\bar{\omega} = 10$.
The oscillators have the phase difference $\alpha$, either 
around $0$ (i.e. in-phase synchronization) or around $\pi$ (anti-synchronization).
The unstable portions of the solutions are plotted as dashed lines.}
\end{figure}

\begin{figure}
\centerline{\psfig{file=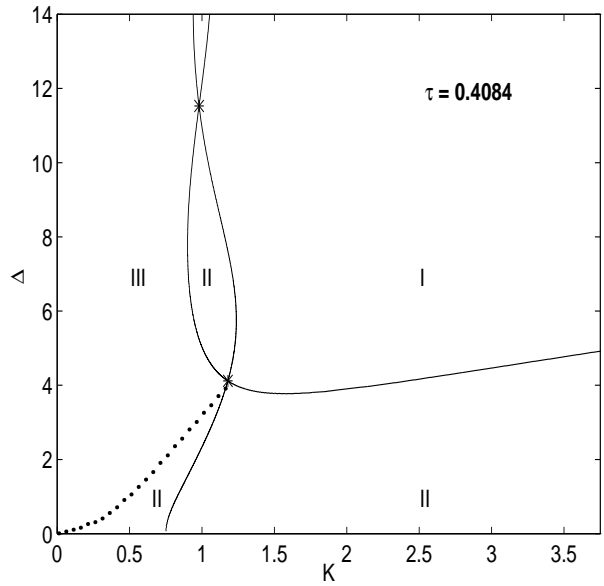,width=8cm,height=8cm}}
\caption{The bifurcation diagram in the ($K,\Delta$) space 
for $\tau = 0.4084$ and 
$\bar{\omega} = 10$. The region marked 
I is the amplitude death region, II corresponds to the phase locked solutions,
and the region III is the phase drift or incoherent region. The dotted curve
which separates the phase locked region from the incoherent region is obtained
by numerical integration of the original equations. Notice that the degenerate 
points (marked by stars) have reappeared for this value of the 
delay parameter $\tau$. At higher values of $K$, other bifurcation curves appear 
indicating higher order frequency states as shown in Fig. 6.}
\end{figure}

\begin{figure}
\centerline{\psfig{file=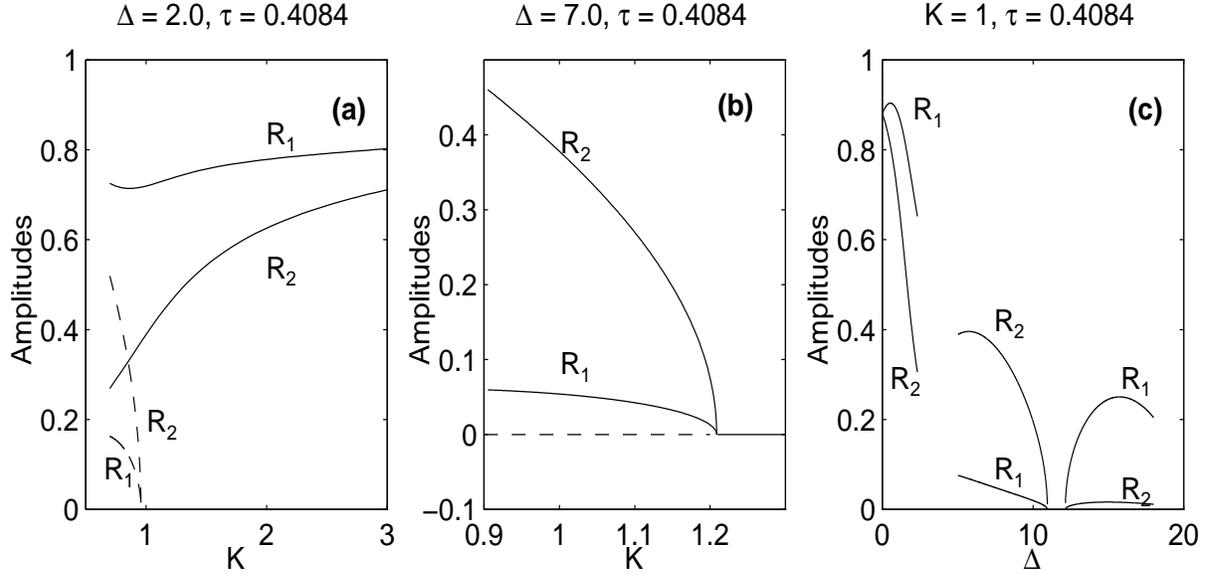,width=16cm,height=8cm}}
\caption{The amplitude curves $R_1$ and $R_2$
of the phase locked solutions of the two
oscillators 
plotted in three different regions. $R_1$ corresponds to the
amplitude of the oscillator with the smaller of the 
intrinsic frequencies and $R_2$ that of the larger of the two.
(a) At $\Delta = 2$, the stable branches (solid curves) continue to exist whereas
the unstable (dashed curves) of the two oscillators merge with the origin as $K$ is
increased. (b) The amplitudes of the oscillators in one of the `closed loops' in the
bifurcation diagram (previous figure) at $\Delta = 7$. The origin is 
unstbale (dashed line) in the region where the periodic orbits are stable
and stable outside this region, illustrating the supercritical Hopf bifurcation.
And (c) shows the phase locked
solutions as one moves along the line $K = 1$. In the first gap there are phase drift
solutions and in the second gap amplitude death solutions.}
\end{figure}

\begin{figure}
\centerline{\psfig{file=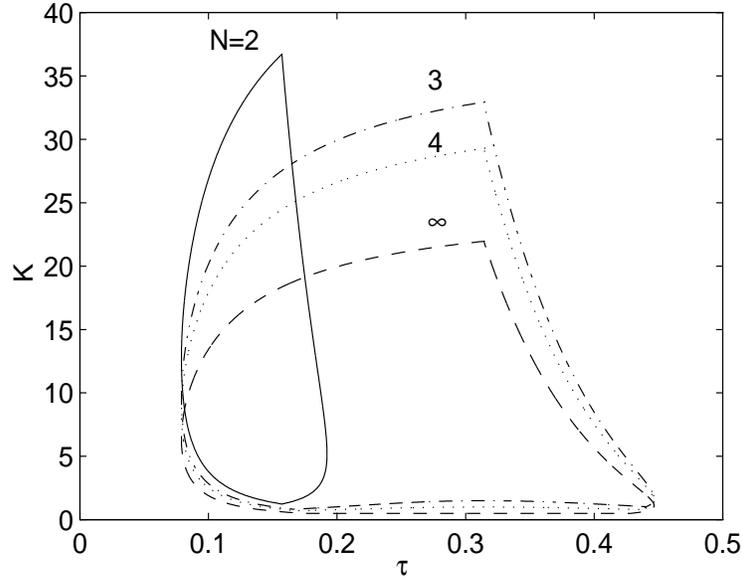,width=10cm,height=8cm}}
\caption{The existence of death islands for an arbitrary number of globally coupled oscillators.
The figure shows the death island regions for $N = 2, 3, 4$ and $\infty$ plotted from 
the Eqs. (\ref{EQN:A}-\ref{EQN:D}) with $n = 0$.
All the oscillators are assumed to have an intrinsic frequency of $\omega_0 = 10$.
The death island survives even in the limit of $N \rightarrow \infty$.}
\end{figure}

\begin{figure}
\centerline{\psfig{file=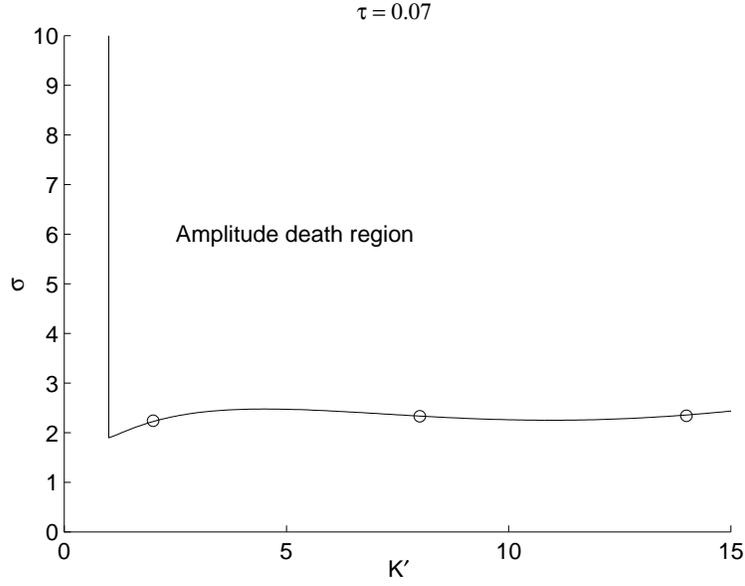,width=10cm,height=8cm}}
\caption{The amplitude death region for N globally coupled oscillators, 
as $N \rightarrow \infty$ for uniform intrinsic distribution of frequencies,
$g(\omega)$, with the average frequency, $\bar{\omega} = 10$. plotted from
the Eq. (\ref{EQN:NDINFtan}) and (\ref{EQN:NDINFlog}) by eliminating $\beta$ 
for $\tau = 0.07$. 
The death region is bounded by $K^\prime = 1$ and 
$\tan(\sigma/K^\prime) = \sigma/(K^\prime-1)$ for $\tau = 0$.
Both the curves meet at a degenerate point $(K^\prime,\sigma) =  (1,\pi/2)$. As the delay
parameter $\tau$ is increased the boundary $K^\prime = 1$ (i.e. $K = 0.5$) 
still continues to be one
of the boundaries. But the second curve distorts in a continuous fashion and
at a certain value of $\tau$ it touches the $\sigma = 0$ axis, showing the death
of identical oscillators. The points marked indicate the numerically found value of 
the boundary for $N = 800$.}
\end{figure}

\begin{figure}
\centerline{\psfig{file=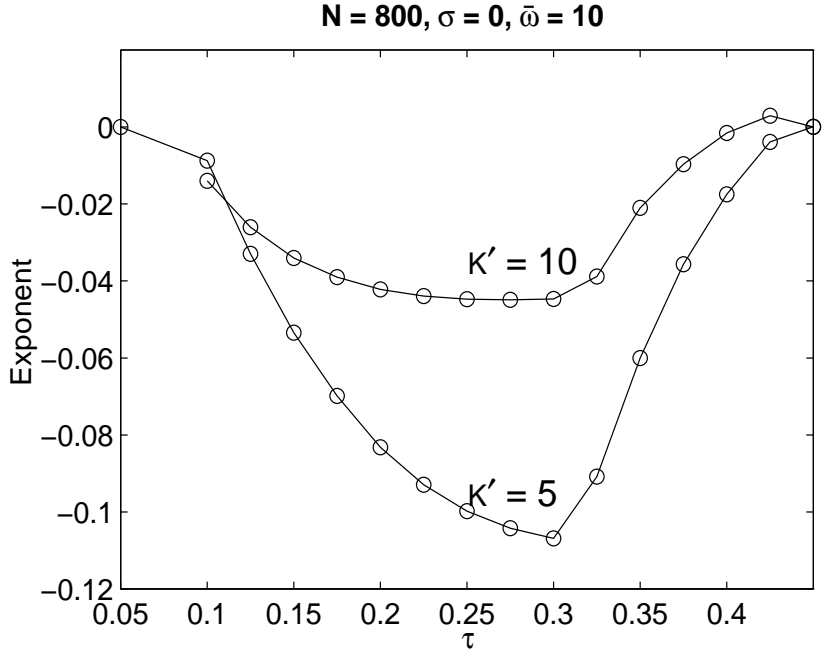,width=11cm,height=9cm}}
\caption{The rate of approach to the origin of the amplitude of the order parameter, R.
If we assume that $R \propto e^{\eta(\tau) t}$, the exponent $\eta(\tau)$ plotted
in the above graph shows that the rate of approach to the origin is maximum
in the middle of the death island. As one approaches the boundaries of the island
on either side the rate decreases to $0$.}
\end{figure}

\begin{figure}
\centerline{\psfig{file=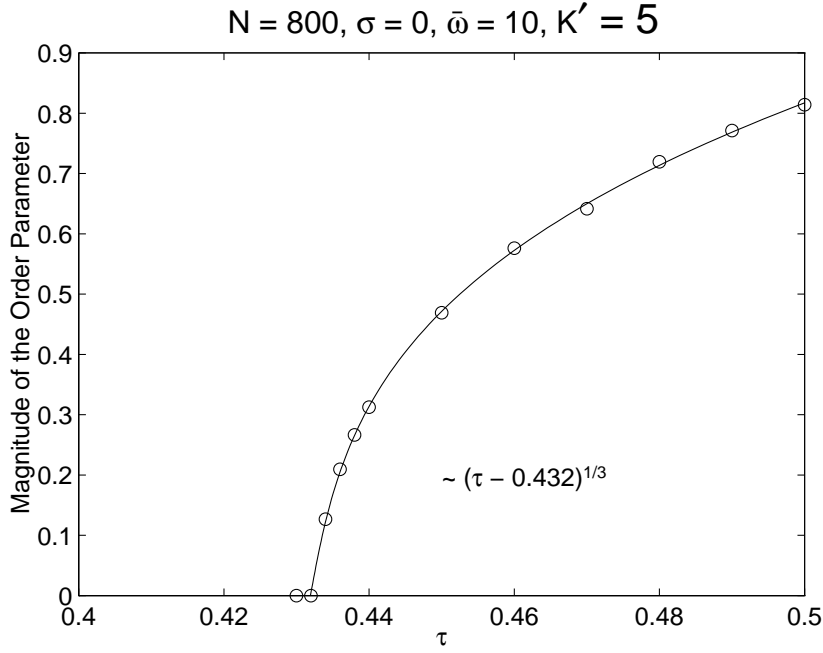,width=11cm,height=9cm}}
\caption{Power law behavior of the magnitude of the order parameter at the 
boundary of the death island.
The order parameter is $0$ inside the death island region due to the fact that
the amplitudes of all the oscillators are zero. The order parameter shows a
power law behavior at the transition between the death island and the phase
locked region. The curve is approximated numerically, at $K^\prime = 5$, by 
$\tau = 0.432 + 0.0992 ( R^3 + 0.022 R^2 + 0.154 R )$.}
\end{figure}

\begin{figure}
\centerline{\psfig{file=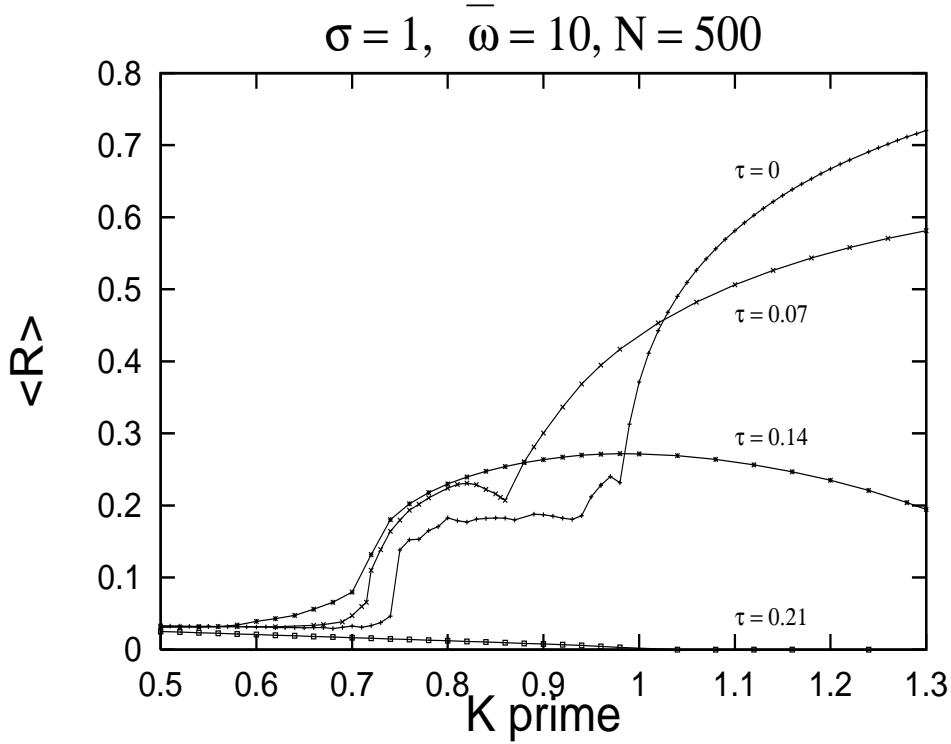,angle=270,width=13cm,height=10cm}}
\caption{The effect of the time delay, $\tau$, on the irregular region which
was found for $\tau = 0$ by Matthews and Strogatz (see text). 
This region
which appears as a plateau for $\tau = 0$ as a whole shrinks as $\tau$ is increased. 
The order parameter is unstable
in this region. However a long temporal record of $R$ gives nearly constant
average value. The average value distinguishes the three distinct regions 
quite accurately. To the right of this plateau region is the phase locked region,
and to the left is the incoherent region.}
\end{figure}

\end{document}